\documentclass[amsmath,amssymb]{rspublic}

\usepackage{graphicx}
\usepackage{dcolumn}
\usepackage{bm}
\usepackage{color}

\begin{document}

\title{Dynamical magnetoelectric effects in multiferroic oxides}
\shorttitle{Dynamical magnetoelectric effects in multiferroic oxides}

\author{Yoshinori Tokura$^{1,2,3}$ and Noriaki Kida$^2$}
\affiliation{$^1$Department of Applied Physics, The University of Tokyo, Tokyo 113-8656, Japan
\\$^2$Multiferroics Project (MF), ERATO, Japan Science and Technology Agency (JST), Tokyo 113-8656, Japan
\\$^3$Cross-Correlated Materials Research Group (CMRG), Advanced Science Institute, RIKEN, Wako 351-0198, Japan
}
\shortauthor{Y. Tokura and N. Kida}

\maketitle

\begin{abstract}
{magnetoelectric effect; multiferroics; domain walls; electromagnons}
Multiferroics with coexistent ferroelectric and magnetic orders can provide an interesting laboratory to test unprecedented magnetoelectric responses and their possible applications. One such example is the dynamical and/or resonant coupling between magnetic and electric dipoles in a solid. As the examples of such dynamical magnetoelectric effects, (1) the multiferroic domain wall dynamics and (2) the electric-dipole active magnetic responses are discussed with the overview of recent experimental observations.
\end{abstract}

\section{Introduction}

Highly efficient control of magnetism in terms of electric field or current in a solid may widen the bottle-neck of the contemporary spin-electronics technology. Since the magnetoelectric (ME) effect, meaning magnetic (electric) induction of polarization $P$ (magnetization $M$), was first confirmed in 1959--60 (Dzyaloshinskii 1960; Astrov 1960), many magnetic materials have been demonstrated to show this effect (Fiebig {\it et al.} 2005). Nevertheless, the magnitude of the observed ME effect has been too small to apply to any practical devices. 

Multiferroics, the materials in which both magnetic and ferroelectric orders can coexist, are the prospective candidate which can potentially host the gigantic ME effect. A naive expectation is that the large ME effect may be driven by weak electric or magnetic field, if the $M$ and $P$ shows the close coupling. There can be several possible ways to materialize the multiferroics. However, the $P-M$ coexistence alone does not necessarily lead to the effective mutual coupling. In this context, the magnetically driven ferroelectricity is the most plausible candidate in which the enhanced $M-P$ coupling is anticipated.

\begin{figure}[bt]
\begin{center}
\includegraphics[width=0.76\textwidth]{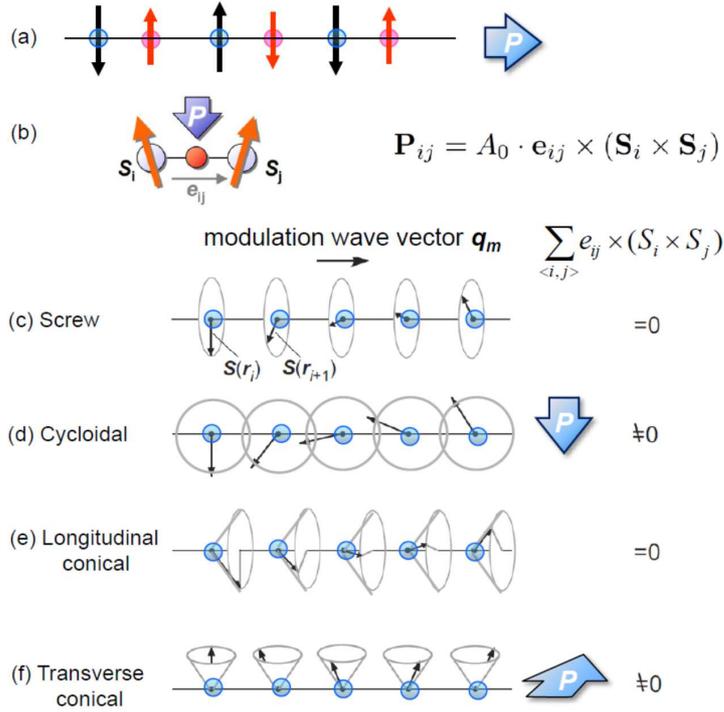}
\end{center}
\caption{(Online version in colour) Inversion symmetry breaking by (a) collinear and (b) noncollinear magnetic order and possible polarization direction. (c)--(f) Schematic illustrations of types of spiral magnetic structures on a one-dimensional array of magnetic moments S(r). They include (c) proper screw, (d) cycloidal, (e) longitudinal-conical, and (f) transverse-conical magnetic structure. The magnitudes of macroscopic polarization calculated from the spin current model or inverse DM model (Katsura {\it et al.} 2003; Sergienko {\it et al.} 2006) are also indicated for their respective structures.}
\label{SS}
\end{figure}

Various types of spin orders can have a potential to break the inversion symmetry and produce the spontaneous (ferroelectric) $P$. This is true even irrespective of its collinear or non-collinear form, when they are placed on some specific lattice geometry. For example, the up-up-down-down collinear spin arrangement along the atomically alternate A-B lattice [Fig. \ref{SS}(a)] can break the inversion symmetry, and in reality the inequivalent interatomic forces (exchange striction) working between the up-up (or down-down) spin pair and the up-down one can produce the $P$. This type of magnetostriction induced ferroelectricity has been found in several materials (Inomata {\it et al.} 1996; Choi {\it et al.} 2008; Tokunaga {\it et al.} 2008); in this case the long-range magnetic order should be commensurate with the lattice periodicity. 

On the other hand, rather lattice-form independent mechanisms for the magnetically induced ferroelectricity have recently been found for versatile compounds with spiral spin orders as depicted in Figs. \ref{SS}(c)--\ref{SS}(f). When the spins on the adjacent atomic sites are mutually canted [Fig. \ref{SS}(b)], the horizontal mirror-plane symmetry is lost, meaning the possible generation of the polarization along the vertical direction (Katsura {\it et al.} 2005; Sergienko {\it et al.} 2006; Mostovoy 2006). Recently, it has been theoretically shown (Katsura {\it et al.} 2005) that the overlap of the electron wave function between the two atomic sites with canted spins generates the genuine electronic polarization via the spin-orbit interaction. In other words, spontaneous spin current flows between the mutually canted spin sites; by analogy to the charge current as producing magnetic field, the spin current produces fictitious electric field or electric polarization.

When the spins form the transverse-spiral (cycloidal) modulation along the specific crystallographic direction [Fig. \ref{SS}(d)], every nearest-neighbor spin pair produces the unidirectional local $P$ and hence the macroscopic $P$ of electronic (magnetic) origin should be generated. The spontaneous polarization can be expressed as
\begin{equation}
P=a\sum_{<i,j>}e_{ij}\times(S_i\times S_j).
\label{SpinCurrent}
\end{equation}
Here, $e_{ij}$ is the unit vector connecting the neighboring spins $S_i$ and $S_j$, and the proportional constant $a$ is determined by the spin-orbit and spin exchange interactions as well as the possible spin-lattice coupling term. The plus/minus of $P$ direction depends on a clock-wise or counter-clock-wise rotation of the spin (called spin helicity) in proceeding along the spiral propagation axis. The model is now termed the spin current model (Katsura-Nagaosa-Balatsky model) on the basis of the above-mentioned duality analogy, or the inverse Dzyaloshinskii-Moriya (DM) model. The latter naming is based on the following scenario (Sergienko {\it et al.} 2006): The conventional DM interaction on the noncentrosymmetric bond causes the canting of the interacting spins. Conversely, the canted spins by some other reasons ($e.g.,$ magnetic frustration) may displace the intervening ion bridging the spin sites so as to generate the new DM vector or the local polarization. In the cycloidal spin order, this inverse DM mechanism can also produce the macroscopic polarization as described in Eq. (\ref{SpinCurrent}).

\section{Perovskite manganites with cycloidal spin orders}

\begin{figure}[bt]
\begin{center}
\includegraphics[width=0.99\textwidth]{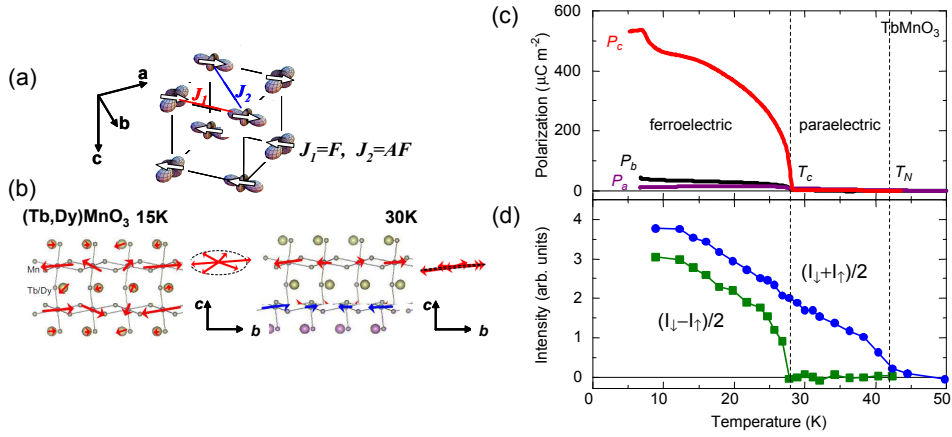}
\end{center}
\caption{(Online version in colour) (a) Schematic illustration of spin/orbital ordered state and magnetic exchange interactions for LaMnO$_3$. (b) The solved magnetic structures at Mn sites in paraelectric antiferromagnetic phase $(T_{\rm c} < T < T_{\rm N})$ and ferroelectric antiferromagnetic phase $(T < T_{\rm c})$ for (Tb, Dy)MnO$_3$ $(q_{\rm m}=1/3)$ (Arima {\it et al.} 2006). (c) Temperature profiles of electric polarization along the $a$, $b$, and $c$ axes for TbMnO$_3$ (Kimura {\it et al.} 2003b). (d) Corresponding temperature dependence of the polarized neutron scattering intensity at $Q$ = (4, $+q_{\rm m}$, 1). $I_{\downarrow}$ and $I_{\uparrow}$ indicate the scattering intensities for the neutron spin parallel and antiparallel to $Q$, respectively (Yamasaki {\it et al.} 2007a). }
\label{RMnO3}
\end{figure}

The strong $M-P$ coupling based on the spiral spin order is prototypically observed for perovskite type manganites $R$MnO$_3$ ($R$ = Tb, Dy, Eu$_{1-x}$Y$_x$) (Kimura {\it et al.} 2003b; 2005; Noda {\it et al.} 2006; Hemberger {\it et al.} 2007; Yamasaki {\it et al.} 2007b) which shows the transverse spiral (cycloidal) spin order, as shown in Fig. \ref{SS}(d). $R$MnO$_3$ is the parent compound of the colossal-magnetoresistance (CMR) oxides (Tokura 2006), but its ground state was thus turned out to be occasionally ferroelectric in nature. Among four $3d$ electrons on a Mn$^{3+}$ site, three are accommodated in the lower-lying $t_{2g}$ orbital and one is in the doubly degenerate $e_{\rm g}$ orbital. The collective Jahn-Teller distortion helps the $e_{\rm g}$-orbital to order in a staggered manner on the $ab$-plane as shown in Fig. \ref{RMnO3}(a). Then, the superexchange interaction between the nearest Mn spins on the $ab$ plane $(J_1)$ is ferromagnetic in nature, while the interaction along the $c$ axis is antiferromagnetic. This is the case typically observed for LaMnO$_3$ with the $A$-type (layered antiferromagnetic) order.  When the ionic size of $R$ site becomes smaller, the orthorhombic (GdFeO$_3$ type) distortion of the perovskite or the Mn-O-Mn bond angle distortion increases and accordingly the in-plane second-nearest antiferromagnetic interaction $(J_2)$ increases to compete with $J_1$. The consequence of such magnetic frustration is the realization of long-period incommensurate magnetic modulation structure (Kimura {\it et al.} 2003a). In the case of TbMnO$_3$ or DyMnO$_3$, the compound first undergoes the collinear sinusoidal antiferromagnetic spin order with the magnetic $q$-vector $q_m=0.25-0.45 b^\ast$ (in the $Pbnm$ setting) with lowering temperature and then subsequently turns to the $bc$-plane cycloidal (transverse-spiral) spin order, as confirmed by neutron diffraction studies (Kenzelmann {\it et al.} 2005; Arima {\it et al.} 2006). The case for the (Tb, Dy)MnO$_3$ with a prescribed commensurate value of $q=1/3$ (Arima {\it et al.} 2006) is shown in  Fig. \ref{RMnO3}(b); the cycloidal order is deformed in reality and well described by the elliptic rotation of spins. In accord with the onset of the cycloidal spin order, the spontaneous polarization $P$ emerges along the $c$ axis. The direction of $P$ is in accord with the prediction of the afore-mentioned spin current model (or inverse DM model) and its plus/minus or equivalently the spin helicity is controlled by the direction of the electric field $E$ during the cooling procedure. Recent first-principles calculations (Malashevich {\it et al.} 2008; Xiang {\it et al.} 2008) with consideration of both electronic and lattice contributions can well reproduce the experimental observation.

The reversal of the spin helicity or vector spin chirality $C (=\sum_i S_i\times S_j)$ upon the $E$-induced reversal of $P$ was experimentally confirmed by measurements of the polarized neutron scattering on TbMnO$_3$ (Yamasaki {\it et al.} 2007a) in which strong asymmetry in polarized neutron scattering intensity at $q=\pm q_m$ is observed.  Figure \ref{RMnO3}(d) shows the temperature dependence of the polarized neutron scattering intensities, $(I_\downarrow+I_\uparrow)/2$ and $(I_\downarrow-I_\uparrow)/2$, for TbMnO$_3$. Here, $I_\downarrow$ and $I_\uparrow$ represent respectively the scattering intensities with the neutron spin parallel and antiparallel to the scattering vector which was set nearly parallel to $a^\ast$ in the experiment. The averaged intensity $(I_\downarrow+I_\uparrow)/2$ shows the onset of the incommensurate spin order $(q_m\sim0.3b^\ast)$ below $T_{\rm N}$, while the differential intensity $(I_\downarrow-I_\uparrow)/2$ coincides with the emergence of the cycloidal (to be exact, transverse-elliptical) spin order below $T_{\rm C}$. At the ferroelectric single-domain state, the differential scattering intensity $(I_\downarrow-I_\uparrow)/2$ is approximately proportional to $m_bm_c$ in the present scattering geometry; here, $m_b$ and $m_c$ is the amplitudes of the $b$ and $c$ components of the $bc$-plane cycloidal spins. Note here that the $m_bm_c$ should be proportional to $P$ in the spin current model or inverse DM model [Eq. (\ref{SpinCurrent})]. The nearly parallel behaviour is observed between the spontaneous $P$ along the $c$ axis [Fig. \ref{RMnO3}(c)] and the differential scattering intensity $(I_\downarrow-I_\uparrow)/2$, which confirms the mechanism of the spin current model or the inverse DM interaction for the present magnetically induced ferroelectricity.

\section{Multiferroic domain walls in perovskite manganites}

\begin{figure}[bt]
\begin{center}
\includegraphics[width=0.99\textwidth]{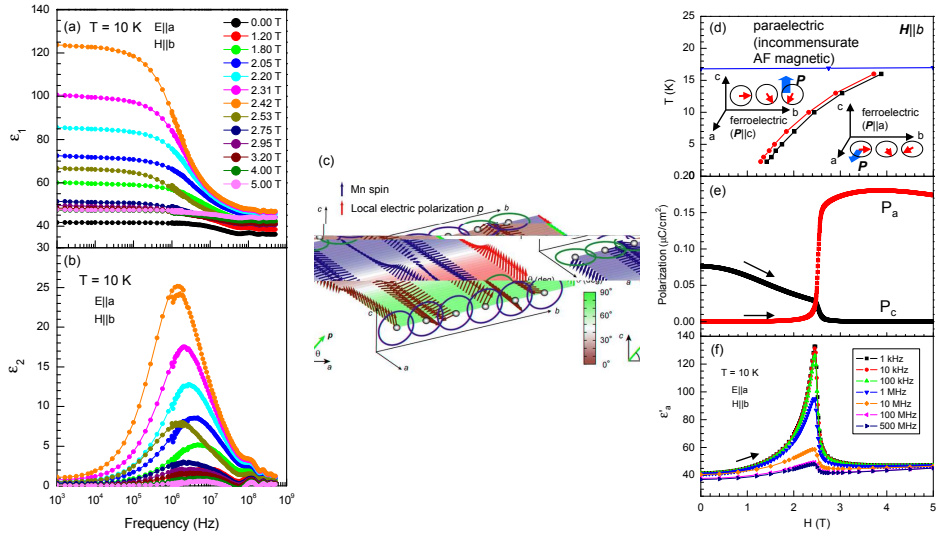}
\end{center}
\caption{(Online version in colour) (a) Real and (b) imaginary part of dielectric constant as a function of frequency measured for DyMnO$_3$ at 10 K under various magnitudes of magnetic field. (c) Calculated domain wall structure between the $P \| +c$ ($bc$-cycloidal) and $P \| +a$ ($ab$-cycloidal) domains. Blue and red arrows represent the Mn spins and local electric polarizations, respectively. The color gradation represents the angle of local electric polarization relative to the $a$-axis. (d) Temperature versus magnetic-field phase diagram for DyMnO$_3$ under $H\| b$ determined from dielectric measurements. $H (\| b)$-dependence of (e) electric polarization ($P \| a$ or $c$) and (f) dielectric constant $(\epsilon\| a)$ are also indicated. Reproduced from Ref. (Kagawa {\it et al.} 2009).}
\label{DyMnO3DW}
\end{figure}

The application of $H$ on the multiferroic $R$MnO$_3$ may often lead the flop of the spin spiral plane, $e.g.,$ from $bc$ to $ab$ plane, which leads the flop of $P$ as well. Such $H$-induced $P$ flop can be viewed as a gigantic magnetoelectric effect, a hallmark of the spiral spin multiferroics. When the $P$ flop is discontinuous with respect to the $H$ magnitude/direction, a new notable magnetoelectric effect is observed due to the coexistence of the different spiral-spin domains with the different spiral planes and/or spiral $q$-vectors. As a prototypical example, we show the magnetoelectric responses for DyMnO$_3$ (Fig. \ref{DyMnO3DW}) (Goto {\it et al.} 2004; Kagawa {\it et al.} 2009) when external $H$ is applied along the $b$ axis; the $P\| c$ or $bc$-plane spin cycloid is turned into the $P\| a$ or $ab$-plane spin cycloid, $e.g.,$ around 2.5 T at 10 K [see Figs. \ref{DyMnO3DW}(d) and \ref{DyMnO3DW}(e)]. The mechanism of the $P$ flop in case of $H\| b$ is rather complex under the competition of the magnetic anisotropy and DM interaction as described in literature (Mochizuki {\it et al.} 2009). The transition is strongly of the first-order nature and hence there may occur the coexistence of the $P\| a$ and $P\| c$ domains. The dielectric constant measured along the $a$-axis shows a large enhancement while scanning $H$ across the phase boundary [Fig. \ref{DyMnO3DW}(f)]. This is one of the largest magnetocapacitance effects observed in multiferroics. The spectra of the dielectric response on the $P$-flop phase boundary (at 2.5 T and at 10 K) are shown in Figs. \ref{DyMnO3DW}(a) and \ref{DyMnO3DW}(b); real and imaginary part of dielectric susceptibility, respectively. The behaviour is typical of Debye type relaxation; the dielectric anomaly upon the $P$-flop transition is thus ascribed not to the resonance type due to some soft mode, but to the lower-lying relaxation mode relevant to the domain wall excitations. Such a dielectric relaxation was suggested in recent literature (Schrettle {\it et al.} 2009) also as the key for magnetocapacitance phenomena.

There are four possible multiferroic domains, $P\|\pm a$ ($ab$-plane spiral with $\pm$spin helicity) and $P\|\pm c$ ($bc$-plane spiral with $\pm$spin helicity), and hence six kinds of domain wall excitations connecting the respective multiferroic domains. Since each polarization direction possesses one-by-one correspondence to a unique set of spin spiral plane and spin helicity, both ferroelectric and magnetic order parameters are expected to change simultaneously across these multiferroic domain walls (MFDW). The present relaxation process, which is present only near the $P$-flop transition region, can be assigned to the dynamical motion of the domain wall between the $P\|\pm a$ and $P\|\pm c$ domains (Kagawa {\it et al.} 2009). To be noted for this MFDW dynamics is that it is dynamically active (up to 1 MHz region) down to low temperature ($e.g.,$ 5 K), sustaining to host the large magnetocapacitance effect. The atomically-thin ferroelectric domain wall in conventional ferroelectrics becomes frozen in dynamics when temperature is decreased from the Curie temperature, in particular at such low temperatures as in the present case. In contrast, the MFDW is magnetic in nature and can be active in analogy to the $H$-driven activity of the ferromagnetic domain wall. The simulation based on the Heisenberg model with suitable parameters for multiferroic DyMnO$_3$ predicts the MFDW running parallel to the $bc$-plane, as shown in Fig. \ref{DyMnO3DW}(c) (Kagawa {\it et al.} 2009); the $bc$-plane and $ac$-plane spin cycloids vary smoothly across the MFDW.

\section{Electromagnons in perovskite manganites}

Another important topic of the dynamical ME effect is the electric-dipole active magnetic resonance, termed electromagnon. Conventionally, the magnetic resonance is driven by the magnetic field $H^\omega$ component of electromagnetic wave or light as appearing in the magnetic permeability spectrum $\mu(\omega)$ in gigahertz to terahertz frequency range. By contrast, the electromagnon is a magnetic {excitation driven by the electric field $E^\omega$ component of light. Therefore, it emerges as the magnetic resonance in the dielectric constant spectrum $\epsilon(\omega)$. Although the electromagnon was theoretically considered since 1970 (Bar'yakhtar {\it et al.} 1970; they originally named it seignetomagnon), the possible signature of electromagnons was recently observed in $\epsilon(\omega)$ of prototypical multiferroics $R$MnO$_3$ by using a backward wave oscillator (BWO) as a light source; at 2.9 meV for TbMnO$_3$ and at 2.5 meV for GdMnO$_3$ around 10 K in zero $H$ ($bc$ spiral spin phase with $P_{\rm s}\| c$) (Pimenov {\it et al.} 2006a). Pimenov {\it et al.} also clarified that the observed absorption becomes electrically active for $E^\omega\| a$ but reduces in intensity upon the magnetic transition from the $bc$ spiral to the $A$-type AFM order by applying the external $H$ along the $c$-axis. Based on these results, they discussed the possible emergence of electromagnons in these compounds. Their pioneering work has stimulated the investigations concerning the low-energy spin dynamics in a variety of multiferroics (Kida {\it et al.} 2009).

Considerable efforts have been devoted to elucidate the microscopic origin of electromagnons in $R$MnO$_3$. On the basis of the spin-current model that explains the ferroelectricity in $R$MnO$_3$, Katsura {\it et al.} discussed the possibility that the electromagnon appears at terahertz frequencies along the direction perpendicular to the spiral spin plane due to the oscillation of the spiral spin plane;  $E^\omega \| a$ and $E^\omega \| c$ in $bc$ and $ab$ spiral spin ordered phases, respectively (Katsura {\it et al.} 2007). Soon after, Senff {\it et al.} performed polarized inelastic neutron scattering experiments on TbMnO$_3$ in the $bc$ spiral spin ordered phase and found the magnetic excitation around 2 meV at $k=0$ magnetic zone-center (Senff {\it et al.} 2007). The observed peak position is nearly identical to that of the electromagnon in TbMnO$_3$ revealed by BWO spectroscopy. 

However, more recent spectroscopic studies on light-polarization and $H$ dependences for DyMnO$_3$ have proven that an origin of electromagnons is neither the modulation of $P_{\rm s}$ nor the rotation mode of the spiral spin plane (Kida {\it et al.} 2008) but perhaps related to the magnetic modulation of the local electric-dipoles via the underlying ionic lattice (Vald\'{e}s Aguilar {\it et al.} 2009) or orbital ordering (Miyahara {\it et al.} 2008). As an example for the basic feature of electromagnons in $R$MnO$_3$, we show in Fig. \ref{EM} the complete set of the light-polarization dependence of the optical spectra $\epsilon\mu(\omega)$ of DyMnO$_3$ in the $bc$ spiral spin ordered phase, as revealed by terahertz time-domain spectroscopy (Kida {\it et al.} 2008). At terahertz frequencies (1 THz $\approx$ 4 meV), there is an antiferromagnetic resonance (AFMR) driven by $H^\omega$. The respective $\epsilon\mu(\omega)$ spectra for the crystal plates ($ac$, $ab$, and $bc$ end-surfaces) with respect to various light-polarizations ($E^\omega$ and $H^\omega$) can isolate the electric and magnetic contributions. Indeed, AFMRs can be seen in $\epsilon\mu(\omega)$ around 2 meV when $H^\omega$ was set parallel to the $a$-axis; we can see the nearly same spectral shape in $\epsilon\mu(\omega)$ for $E^\omega\| c$ and $H^\omega\| a$ [Fig. \ref{EM}(d)] and $E^\omega\| b$ and $H^\omega\| a$ [Fig. \ref{EM}(e)] in spite of different $E^\omega$ polarizations. On the other hand, the pronounced broad absorption in $Im[\epsilon\mu(\omega)]$ composed of two peak structures around 2 meV and 6 meV is discerned for $E^\omega\| a$ and $H^\omega\| c$ [Fig. \ref{EM}(a)]; the lower peak energy nearly coincides with that of AFMR for $H^\omega\| a$. In accord with this absorption, there is a dispersive structure in $Re[\epsilon\mu(\omega)]$. Its magnitude reaches the maximum (about 30), twice as large as that of $Re[\epsilon\mu(\omega)]$ $(= 15-17)$ in other geometries [Figs. \ref{EM}(c), (d), (e), and (f)]. Such a broad absorption is clearly identified only for $E^\omega\| a$; the nearly same spectral shape including the peak position and the magnitude is also observed for $E^\omega\| a$ and $H^\omega\| b$ in spite of different $H^\omega$ polarizations [Fig. \ref{EM}(b)]. In addition, this absorption grows in intensity with lowering temperature below $T_{\rm N}=42$ K and is further enhanced below $T_{\rm c}=19$ K, as shown in the temperature dependence of $\epsilon\mu(\omega)$ spectra for $E^\omega\| a$ and $H^\omega\| c$ [Fig. \ref{EM}(g)]. Therefore, the observed gigantic broad absorption band expanding in the energy region 1--10 meV can be ascribed to the electromagnon for $E^\omega\| a$.

\begin{figure}[bt]
\begin{center}
\includegraphics[width=0.99\textwidth]{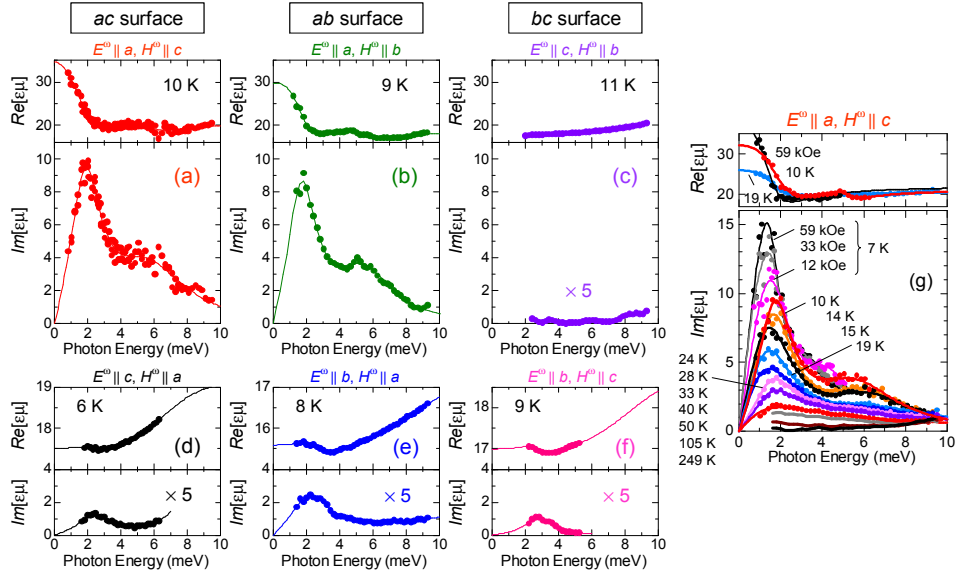}
\end{center}
\caption{(Online version in colour) (a)--(f): Real (upper panels) and imaginary (lower panels) parts of $\epsilon\mu$ spectra of DyMnO$_3$ with respect to $E^\omega$ and $H^\omega$ using crystal plates of $ac$, $ab$, and $bc$, measured around 10 K. (g) $\epsilon\mu$ spectra for $E^\omega\| a$ and $H^\omega\| c$ with varying temperature and magnetic field. Reproduced from (Kida {\it et al.} 2008; 2009).}
\label{EM}
\end{figure}

To clarify the dominant role of the spiral spin order in electromagnons of $R$MnO$_3$, the effect of external $H$ on the $\epsilon\mu(\omega)$ spectra was studied in DyMnO$_3$ for $E^\omega\| a$ and $H^\omega\| c$. At 7 K, the direction of the spiral spin plane is flopped at 20 kOe from $bc$ to $ab$ by applying $H$ along the $b$-axis. The measured $\epsilon\mu(\omega)$ spectra are also included in Fig. \ref{EM}(g). Contrary to the expected characteristic of the rotation mode of the spiral spin plane, that is, the change of the polarization of the electromagnon absorption from $E^\omega\| a$ to $E^\omega\| c$, the almost identical electromagnon absorption is discerned with $E^\omega\| a$ even for the $ab$ spiral spin order, as exemplified by $\epsilon\mu(\omega)$ in $H$ of 33 kOe and 59 kOe. This is a direct proof that the electromagnon is only active along the $a$-axis, irrespective of the direction of the spiral spin plane. Lately, the detailed light-polarization dependence of $\epsilon\mu(\omega)$ was measured in the $ab$ spiral spin phase of Gd$_{0.7}$Tb$_{0.3}$MnO$_3$, which also indicates that the electromagnon only appears along the $a$-axis (Kida {\it et al.} 2009).

One plausible scenario to explain the unique selection-rule of the electromagnon in $R$MnO$_3$ is to consider the symmetric exchange interaction between the mutually canted spins on the underlying perovskite lattice (Vald\'{e}s Aguilar {\it et al.} 2009; Miyahara {\it et al.} 2008), in which local electric dipoles are built in via the Mn-O-Mn bond distribution or staggered $e_g$ orbital order. The calculation based on the cycloidal spin structures in the orbital ordered state can explain the basic feature of the higher-lying electromagnon around 8 meV in terms of the electric-dipole active zone-edge magnon. The validity of this assignment was confirmed by the systematic investigations on the electromagnon spectra of $R$MnO$_3$ with varying the ionic radius of $R$ or equivalently the exchange energy $(J)$ (Lee {\it et al.} 2009). However, there is still an open question concerning the origin of the lower-lying electromagnons around 2 meV, whose energy is nearly identical to the peak position of $k=0$ AFMRs (zone-center magnon). Another important feature of the electromagnon is a coupling with optical phonons; the spectral weight of the electromagnon is transferred from that of the lowest phonon mode along the $a$-axis (Pimenov {\it et al.} 2006b; Vald\'{e}s Aguilar {\it et al.} 2007; Takahashi {\it et al.} 2008). On the contrary, there is no remarkable anomaly in the phonon dispersion curve along the $c$-axis at $T_{\rm N}$ and $T_{\rm c}$, as revealed by inelastic X-ray scattering experiments on TbMnO$_3$ (Kajimoto {\it et al.} 2009).

\section{Dynamical magnetoelectric effects in conical spin ferroelectrics }

Up to now, the inverse DM model (Sergienko {\it et al.} 2006) or the spin-current model (Katsura {\it et al.} 2005) as expressed by Eq. (\ref{SpinCurrent}) has been quite useful to find ample examples of multiferroelectricity. Furthermore, this strategy can be applied also to the transverse-conical spin state [Fig. \ref{SS}(f)] in which the ferromagnetic (homogeneous) and spiral components of the magnetization can coexist. Note that the transverse-conical order can also give rise to the electric polarization along the direction parallel to the cycloidal plane and perpendicular to the $q$-vector, according to Eq. (\ref{SpinCurrent}). This may lead to the true ($i.e.,$ in the sense of both ferromagnetic and ferroelectric) multiferroics of magnetic origin. Here we present the two prototypical cases; spinel type CoCr$_2$O$_4$ (Yamsaki {\it et al.} 2006; Choi {\it et al.} 2009) and Y-type hexaferrite Ba$_2$Mg$_2$Fe$_{12}$O$_{22}$ (Ishiwata {\it et al.} 2008). The former is a generic transverse-conical magnet [Fig. \ref{SS}(f)], while the latter is an originally longitudinal conical magnet [Fig. \ref{SS}(e)] but turned into a transverse state with a low magnetic field applied along the slanted direction off the screw axis ($q$-vector).

The spinel type CoCr$_2$O$_4$ (Fig. \ref{CoCr2O4}) undergoes the ferrimagnetic transition at $T_{\rm C}=93$ K, and with further lowering temperature the transition to the transverse-conical spin state, $i.e.,$ the ferromagnetic plus transverse-spiral spin state, with the incommensurate propagation vector of ($q$, $q$, 0) $(q\sim0.63)$ takes place at $T_{\rm S}=26$ K (Menyuk {\it et al.} 1964; Lyons {\it et al.} 1962; Tomiyasu {\it et al.} 2004). The incommensurate-commensurate or lock-in transition occurs around 15 K without changing much the $q$-value (Funahashi {\it et al.} 1987). As shown in the left panel of Fig. \ref{CoCr2O4}, the spiral plane on which the rotating component of spin lies is the (001) plane, while the spontaneous magnetization $(M)$ directs along the [001] or equivalent directions. Then, according to the spin current model [Eq. (\ref{SpinCurrent})], the spontaneous polarization vector is expected to lie along the axis perpendicular to the [001] $M$ direction (see the left panel of Fig. \ref{CoCr2O4}).

\begin{figure}[bt]
\begin{center}
\includegraphics[width=0.99\textwidth]{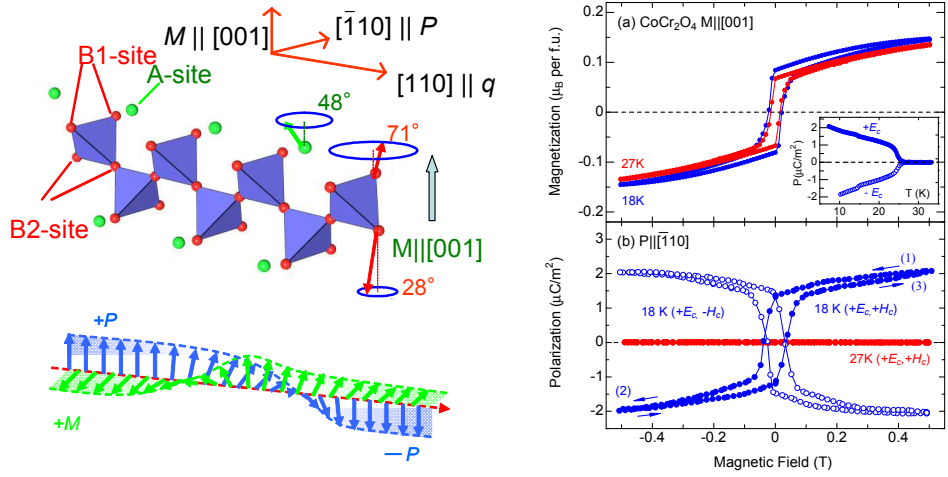}
\end{center}
\caption{(Online version in colour) Magnetic field dependence of (a) magnetization and (b) electric polarization measured for CoCr$_2$O$_4$ above (27 K) and below (18 K) the ferroelectric transition temperature ($T_{\rm s} = 26$ K). For measurements of polarization, the magnetic field was scanned between $+H_{\rm c}$ and $-H_{\rm c}$, for each magnetoelectrically-cooled state prepared with $(E_{\rm c}, H_{\rm c})$ and $(E_{\rm c}, -H_{\rm c})$, as represented by closed and open circles. $E_{\rm c}$ (=400 kVm$^{-1}$) and $H_{\rm c}$ (=0.5 T) stand for the cooling electric and magnetic fields, respectively. The inset to (a) shows the temperature dependence of spontaneous polarization in the cases of the magnetoelectric cooling with positive and negative $E_{\rm c}$. In the left panels, the crystal and magnetic structure of CoCr$_2$O$_4$ and the plausible clamping process of the ferromagnetic and polarization domain walls are indicated (see text). Reproduced from Ref. (Yamasaki {\it et al.} 2006).}
\label{CoCr2O4}
\end{figure}

The inset to Fig. \ref{CoCr2O4}(a) shows the temperature dependence of polarization $(P)$, which was deduced from the pyroelectric current in zero electric- and zero magnetic-field warming after the  magnetoelectric (ME) cooling procedure (Yamasaki {\it et al.} 2006). The $P$ shows the onset at $T_{\rm s}$ ($=26$ K). The sign of the $P$ was confirmed to coincide with that of the poling electric field $E_{\rm c}$ applied in the ME cooling procedure. Thus, the observed $P$ is the spontaneous polarization in the ferroelectric phase below $T_{\rm S}$, irrespective of the spin spiral state being commensurate or incommensurate (The tiny anomaly of temperature variation of $P$ is due to the lock-in transition of the conical state at $T_{\rm C-IC}=15$ K).

We show in Fig. \ref{CoCr2O4}(b) the magnetic field dependence of $P$ at selected temperatures (18 K and 27 K) in comparison with the conventional magnetization curve [Fig. \ref{CoCr2O4}(a)]. Prior to the magnetic field scan, the $P$ direction could be completely determined by the direction of the cooling electric field ($E_c=400$ kV/m) in the ME cooling procedure ($H_{\rm c}= 0.5$ T), as shown in the inset. Then the magnetic field was scanned between $+H_{\rm c}$ and $-H_{\rm c}$. As clearly seen in Fig. \ref{CoCr2O4}(b), whichever direction of the magnetization is taken as the starting point, the $P$ is always reversed upon the reversal of the magnetization direction [Fig. \ref{CoCr2O4}(a)].

Reflecting this $M-P$ coupling, the synchronized reversal of the spontaneous polarization and magnetization can be confirmed by the sequential scan of the magnetic field between $+0.2$ T and $-0.2$ T, $e.g.,$ at 18 K (not shown) (Yamasaki {\it et al.} 2006). The magnitude of flopped $P$ is almost constant, retaining the full value as observed by the quasi-static measurement. The ferroelectric and ferromagnetic single-domain feature, which was achieved by the ME cooling procedure, is almost perfectly maintained during the magnetic-field induced reversal of the magnetization and the polarization.  In other words, the ferromagnetic and ferroelectric domain walls are anticipated to be always clamped, as schematically depicted in the left panel of Fig. \ref{CoCr2O4}:  The magnetic domain wall, $e.g.,$ of the Bloch type, may sustain the similar spin-spiral habit and hence the direction of the $P$ may also rotate, while keeping the orthogonal relation with $M$.

\begin{figure}[bt]
\begin{center}
\includegraphics[width=0.99\textwidth]{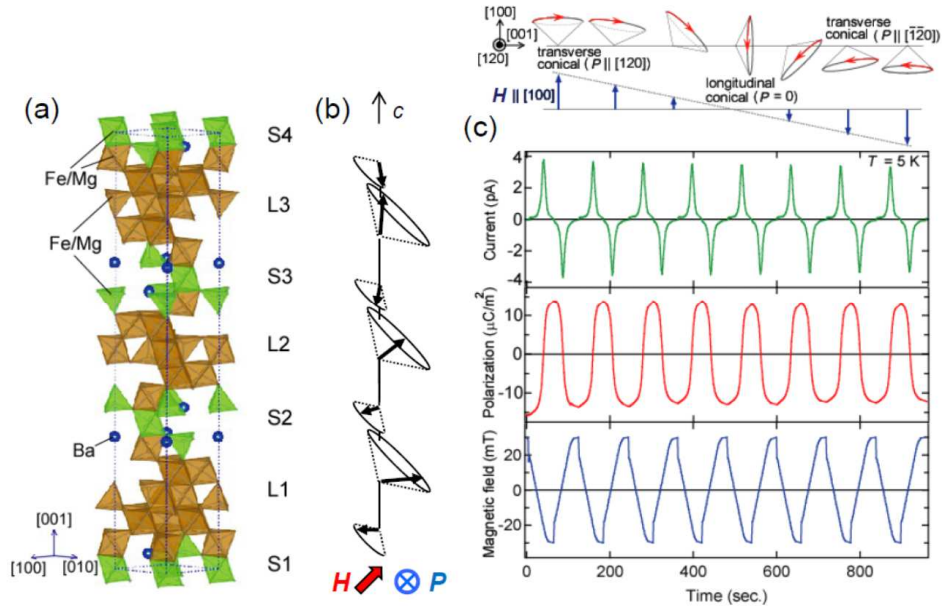}
\end{center}
\caption{(Online version in colour) (a) Schematic crystal structure of Ba$_2$Mg$_2$Fe$_{12}$O$_{22}$. The magnetic structure consists of alternate stacks of L-blocks (brown) and S-blocks (green) having large and small magnetic moments, respectively. (b) Illustration of slanted conical spin structure at $T < 195$ K and $\mu_0 H\sim30$ mT. (c) Oscillating polarization current, cyclic reversal of ferroelectric polarization, against periodically changing magnetic field $(H)$ applied parallel to [100], plotted as a function of time. The drawing at the top schematically shows the temporal change of the spin cone and helicity while oscillating $H$. Reproduced from Ref. (Ishiwata {\it et al.} 2008).}
\label{BMFO}
\end{figure}

Another good example of the conical spin multiferroics is the Y-type hexaferrites, whose structure is schematically shown in Fig. \ref{BMFO}(a). Ba$_2$Mg$_2$Fe$_{12}$O$_{22}$ (BMFO) undergoes the spin-collinear ferrimagnetic transition at 553 K, and then the proper screw spin transition at 195 K, according to the neutron diffraction studies by Momozawa {\it et al.} (Momozawa {\it et al.} 1986). The Fe spins collinearly order within each structural block, such as S and L blocks shown in Fig. \ref{BMFO}(a), but show every 70$^\circ$ rotation in propagating along the $c$ axis. With further lowering temperature below 50 K, the proper screw structure turns into the longitudinal conical structure [Fig. \ref{SS}(e)], exhibiting the ferromagnetic component along the $c$-axis. According to the spin current model, this longitudinal conical spin structure cannot produce the spontaneous $P$, while the local $P$ should show the proper screw like modulation along the $q$-vector $(\| c)$. By applying the external magnetic field to cant the cone axis off the original $c$ axis or $q$ direction, the canted-conical spin state can generate the $P$ normal to both the magnetic $q$-vector $(\| c)$ and the cone axis $\| H$, and hence directing the in-plane direction, as shown in Fig. \ref{BMFO}(b). As the extreme situation for this, when the $H$ is applied normal to the $c$ axis but small enough ($< 0.1$ T in the present case), the simple transverse-conical structure [Fig. \ref{SS}(f)] is realized while maintaining the incommensurate $q$-vector along the $c$ axis. Figure 6(c) shows the synchronous change of the polarization current and polarization when the in-plane $H$ as low as 30 mT is repeatedly reversed (Ishiwata {\it et al.} 2008). The endurable feature of the magnetically induced polarization vector indicates the continuous rotation of the spin cone with the reversal of the in-plane $H$, as schematically shown in the upper panel of Fig. \ref{BMFO}(c). It is worth noting here that a sort of magneto-electric induction as observed in this case can be alternatively interpreted in terms of the spin current mechanism. As described above, the mutually canted spins can generate the spontaneous spin current between the spin sites, and in the present case along the $c$ axis. The temporal variation of this spin current can generate the inverse spin Hall effect (Nagaosa 2008), that is nothing but the ac polarization current as observed. 

When the transversely applied $H$ exceeds some critical value, $e.g.,$ around 0.3 T, in the conical state below 50 K, BMFO undergoes the commensurate order with the periodicity of the four (S, L) block units (Ishiwata {\it et al.} 2009). This magnetic state can generate the ferroelectric state with a larger $P$ value ($\sim100$ $\mu$C/m$^2$), although the state is distinct from the above-described lower-$H$ phase. The high-$H$ ferroelectric phase was perhaps identical with that reported for the related hexaferrite Ba$_{0.5}$Sr$_{1.5}$Zn$_2$Fe$_{12}$O$_{22}$ by Kimura {\it et al.} (Kimura {\it et al.} 2005b) as the possible high-temperature multiferroic, and appears to be generic for the Y-type hexaferrites with proper-screw or longitudinal-conical spin order in zero field. According to recent studies (Ishiwata {\it et al.} 2009), this high-$H$ multiferroic state shows the commensurate transverse-conical state, hence shares the common microscopic origin of the $P$ with the lower-$H$ incommensurate state.

\section{Concluding remarks}

For the multiferroics with transverse-spiral spin orders, we have argued two kinds of  dynamical magnetoelectric (ME) effects; the dynamics of multiferroic domain wall (MFDW) and the electromagnon or electrically-driven magnetic resonance. To enhance the ME response on these multiferroics, the consideration of the MFDWs and their dynamics is important. There are several types of multiferroic domain walls (MFDWs) with respect to (1) spin vector chirality and (2) spin spiral $q$-vector. Here, we have shown the case of the mobile MFDWs between the orthogonal spiral plane (vector spin chirality) domains which are nearly degenerate. The magnetic-field tuning of the degeneracy for the case of DyMnO$_3$ with $bc$- or $ab$-plane spin-spiral plane leads to enhancement of the dielectric response due to the movement of MFDW. The MFDW is anticipated to take a shape of the complex curved spin texture comprising the crossover from the $bc$- to $ab$-spiral state. Such a MFDW response can lead also to the unfrozen gigantic magnetocpacitance effect down to low temperatures. Another interesting case is the MDFW bridging over the different-$q$ domains in the high-symmetry ($e.g.,$ triangular-lattice) multiferroics (Seki {\it et al.} 2009).

For the transverse-conical magnet, in which the spontaneous magnetization $M$ can coexist with the induced electric polarization $P$, the clamping of $M$ and $P$ on the domain wall is most important toward the $E$-control of the $M$ domains. In conical magnet CoCr$_2$O$_4$, the low-temperature multiferroic state shows the $P$ reversal upon the $M$ reversal, indicating the strong clamping, although the $P$ value in this system is too small to achieve the $E$-control of the $M$ domain. It has recently been demonstrated that the $E$-control of the weak-ferromagnetic domains in GdFeO$_3$ is partly possible via the $M-P$ clamped MFDW while the two other composite domain walls show no $P-M$ clamping (Tokunaga {\it et al.} 2009). On the other hand, the flexible $P$-direction control with rotating $H$ is possible via the rotation of the spin cones for many cycloidal or conical magnets.

The electromagnon excitation is viewed as the resonant $E$-control of spins. Here, the activation mechanism of the electromagnon is either the $E$-modulation of the exchange interaction of symmetric type or antisymmetric type. As demonstrated for the electromagnon for $R$MnO$_3$ whose multiferroicity originates from the antisymmetric exchange interaction (spin-current or inverse DM model), the symmetric exchange interaction between the noncollinear spins are the likely origin of the $E$-activity. However, we may conversely anticipate the electromagnon excitation in spin-collinear magnets via the $E$-action on the antisymmetric exchange interaction. The ubiquity of the electromagnons in existing versatile magnets is quite likely; this should be a big challenge also in the light of the ultrafast $E$-control of the interacting spins.

\end{document}